\documentclass[seceq,preprint]{ptptex}

\newtheorem{theorem}{Theorem}[section]
\def\endproof{\hskip0.6em plus0.1em minus0.1em
\setbox0=\null\ht0=5.4pt\dp0=1pt\wd0=5.3pt
\vbox{\hrule height0.8pt
\hbox{\vrule width0.8pt\box0\vrule width0.8pt}
\hrule height0.8pt}}

\newcommand{\tr}{\mathop{\rm tr}\nolimits}
\newcommand{\rmd}{{\rm d}}

\preprintnumber[3cm]{%<-- [..]: optional width of preprint # column.
IU-MSTP/61\\{\tt hep-lat/0407010}}
\title{%        %You can use \\ for explicit line-break
A No-go Theorem for the Majorana Fermion on a Lattice%
}

\author{%       %Use \scshape  for the family name
Hiroshi \textsc{Suzuki}\footnote{E-mail: hsuzuki@mx.ibaraki.ac.jp}
}

\inst{%         %Affiliation, neglected when [addenda] or [errata]
Institute of Applied Beam Science, Ibaraki University, Mito 310-8512, Japan
}

\abst{%         %this abstract is neglected when [addenda] or [errata]
A variant of the Nielsen--Ninomiya no-go theorem is formulated. This theorem
states that, under several assumptions, it is impossible to write down a
doubler-free Euclidean lattice action of a single Majorana fermion in $8k$ and
$8k+1$ dimensions.
}

\begin{document}

\maketitle

\section{Introduction}
In Ref.~\citen{Inagaki:2004ar}, the authors (including the present author)
examined compatibility of the Majorana decomposition and the charge
conjugation property of lattice Dirac operators by considering the
Wilson--Dirac operator\cite{Wilson:1975id} and the overlap-Dirac
operator\cite{Neuberger:1998fp} as examples. There, it was observed that the
Majorana decomposition cannot be applied in $8k$ and $8k+1$ dimensions and, as
a consequence, it is impossible to obtain a physically acceptable lattice
formulation of the Majorana fermion in these dimensions. The authors then
argued that this difficulty associated with the Majorana fermion is a
manifestation of the global gauge anomaly\cite{Witten:fp} in
$8k$~dimensions.\cite{Holman:ef} If this argument is correct, it will be
extremely difficult to realize a physically acceptable lattice formulation of
the Majorana fermion in these dimensions, because the difficulty has an
intrinsic physical meaning.

In the present paper, we formulate a variant of the Nielsen--Ninomiya no-go
theorem\cite{Nielsen:1980rz,Friedan:nk} which states, under several
assumptions, the impossibility of a lattice Majorana fermion in $8k$ and $8k+1$
dimensions. Our result, which is independent of the particular choice of the
lattice Dirac operator and reflects the characteristics of the Clifford
algebra in these dimensions, provides further evidence that the difficulty
associated to the Majorana fermion in these dimensions has a deeper origin. In
addition, we believe that this theorem will be useful in formulating a method
to avoid the difficulty, because our theorem precisely specifies the
assumptions that lead to the difficulty. One of these assumptions must be
removed in order to realize a lattice formulation of the Majorana fermion, just
as one usually circumvents the Nielsen--Ninomiya theorem by removing the
assumption on chiral symmetry.

\section{Basic assumptions and the theorem}

We consider a single free Majorana fermion defined on a Euclidean lattice of
infinite extent. For this, we assume the following bi-linear form of the
lattice action of the Majorana fermion:
\begin{equation}
   S_{\rm M}=a^d\sum_{x}a^d\sum_{y}{1\over2}\chi(x)^TBD(x-y)\chi(y).
\label{twoxone}
\end{equation}
Here  $d$ is the dimensionality of the lattice, with $d=8k$ or~$d=8k+1$, and
the field~$\chi(x)$, which represents the Majorana degrees of freedom, is
Grassmann odd. The matrix~$B$ represents the ``charge conjugation matrix''
satisfying the relations
\begin{eqnarray}
   &&B\gamma_\mu B^{-1}=(-1)^n\gamma_\mu^*=(-1)^n\gamma_\mu^T,
\\
   &&B^{-1}=B^\dagger,\qquad B^T=(-1)^{n(n+1)/2}B,\qquad
   B^*B=(-1)^{n(n+1)/2},
\end{eqnarray}
with $n=[d/2]$. (This is the matrix~$B_1$ of Ref.~\citen{Inagaki:2004ar}.) For
even $d=2n$, we also define the chiral matrix~$\gamma$ by
\begin{eqnarray}
   &&\gamma=i^{-n}\gamma_0\gamma_1\cdots\gamma_{2n-1},\qquad
   \{\gamma,\gamma_\mu\}=0,\qquad\gamma^\dagger=\gamma,\qquad\gamma^2=1,
\\
   &&B\gamma B^{-1}=(-1)^n\gamma^*=(-1)^n\gamma^T.
\end{eqnarray}
We have assumed that the lattice Dirac operator~$D(x,y)$ possesses
translational invariance; i.e, it depends only on the difference of positions,
$x-y$. We also assume that the Dirac operator is local; the precise definition
of this property is given below. To reproduce the Euclidean action of the
Majorana fermion in the continuum theory in the classical continuum limit, it
is necessary to assume
\begin{equation}
   D(z)=\sum_\mu\gamma_\mu\partial_\mu\delta^d(z)+O(a).
\end{equation}
See Ref.~\citen{Inagaki:2004ar} for background discussion concerning the above
construction, the so-called ``Majorana decomposition.''

Now, because the field~$\chi(x)$ is Grassmann odd, the Dirac operator
in~Eq.~(\ref{twoxone}) must be skew-symmetric, i.e., $D(-z)^TB^T=-BD(z)$,
or\footnote{In this paper, it is understood that the transpose operation acts
only on the spinor indices. The transpose operation in
Ref.~\citen{Inagaki:2004ar}, $D^T$, which acts also on position-space indices,
corresponds to~$D(-z)^T$ in the present notation.}
\begin{equation}
   D(-z)^T=-BD(z)B^{-1}.
\label{twoxseven}
\end{equation}
According to the Nielsen--Ninomiya theorem, the Dirac operator~$D$ cannot
possess chiral invariance because otherwise species doubling occurs under
physically reasonable assumptions. Thus we have to assume $\{\gamma,D\}\neq0$.
As the possible nature of this breaking of the chiral symmetry, we postulate
\begin{equation}
   \{\gamma,D(-z)\}^T=+B\{\gamma,D(z)\}B^{-1},
\label{twoxeight}
\end{equation}
in addition to the fundamental requirement~(\ref{twoxseven}). We refer to the
property~(\ref{twoxeight}) as ``pseudo-chiral invariance.'' The postulation of
this property is motivated by the fact that the Wilson--Dirac operator and the
overlap-Dirac operator in $8k$ dimensions satisfy the
relation~(\ref{twoxeight}), because these Dirac operators behave in accordance
with the relation~$D(-z)^T=+B\gamma D(z)\gamma B^{-1}$.

We immediately find, however, that if we require {\it both\/} the
skewness~(\ref{twoxseven}) and the pseudo-chiral invariance~(\ref{twoxeight}),
the Dirac operator~$D$ must possess chiral invariance, i.e., $\{\gamma,D\}=0$.
In other words, for a Dirac operator~$D$ that possesses pseudo-chiral
invariance, skewness can be enforced by ``anti-symmetrizing'' the operator as
\begin{equation}
   D_A(z)={1\over2}[D(z)-B^{-1}D(-z)^TB].
\label{twoxnine}
\end{equation}
Then we have $D_A(-z)^T=-BD_A(z)B^{-1}$ and $\{\gamma,D_A\}=0$. (The latter
relation represents a special case of pseudo-chiral invariance.) For example,
the anti-symmetrization of the Wilson--Dirac operator removes the Wilson term
and makes it chiral invariant, as observed in Ref.~\citen{Inagaki:2004ar}.
It can also be verified that, with the assumption of the pseudo-chiral
invariance~(\ref{twoxeight}), the lattice action~(\ref{twoxone}) possesses the
chiral invariance, i.e., $\delta S_{\rm M}=0$ for
$\delta\chi=\epsilon\gamma\chi$. Thus, recalling the Nielsen--Ninomiya theorem,
we expect species doubling. The precise statement of the situation here is
given by the following.
\begin{theorem}
For the momentum representation of the free lattice Dirac operator
\begin{equation}
   \widetilde D(p)=a^d\sum_ze^{-ipz}D(z)
\end{equation}
in $8k$~dimensions, the following five properties are incompatible:
\begin{enumerate}
\item Locality: $\widetilde D(p)$ is a smooth function of $p_\mu$ with the
period~$2\pi/a$.
\item Correct dispersion: $\widetilde D(p)=i\sum_\mu\gamma_\mu p_\mu+O(ap^2)$.
\item No species doubling: $\widetilde D(p)$ is invertible for all $p\neq0$.
\item Skewness: $\widetilde D(-p)^T=-B\widetilde D(p)B^{-1}$.
\item Pseudo-chiral invariance:
$\{\gamma,\widetilde D(-p)\}^T=+B\{\gamma,\widetilde D(p)\}B^{-1}$.
\end{enumerate}
\end{theorem}

The proof of the theorem is made trivial by invoking the Nielsen--Ninomiya
theorem. From properties 4 and~5, we have $\{\gamma,\widetilde D(p)\}=0$.
This chiral invariance is incompatible with the rest of the properties, 1, 2
and~3, according to the Nielsen--Ninomiya theorem.~\endproof

For odd dimensions, $d=8k+1$, if the Dirac operator~$D$ possesses the ``parity
invariance'' defined by $D(-z)=-D(z)$ or~$D(z)+D(-z)=0$, we immediately
encounter species doubling, as seen below. Thus, we instead postulate the
following ``pseudo-parity invariance'';
\begin{equation}
   [D(z)+D(-z)]^T=+B[D(z)+D(-z)]B^{-1}.
\label{twoxeleven}
\end{equation}
In $8k+1$~dimensions, the Wilson--Dirac operator and the overlap-Dirac operator
possess this property, because these Dirac operators satisfy
$D(z)^T=+BD(z)B^{-1}$. Requiring both the skewness~(\ref{twoxseven}) and the
pseudo-parity invariance~(\ref{twoxeleven}), however, the parity invariance
$D(-z)=-D(z)$ results. In other words, for a Dirac operator~$D$ that
possesses the pseudo-parity invariance~(\ref{twoxeleven}), skewness can always
be enforced by use of Eq.~(\ref{twoxnine}). Then, the anti-symmetrized Dirac
operator satisfies $D_A(-z)=-D_A(z)$. These arguments are summarized as the
following.
\begin{theorem}
For the momentum representation of the free lattice Dirac operator
\begin{equation}
   \widetilde D(p)=a^d\sum_ze^{-ipz}D(z),
\end{equation}
in $8k+1$~dimensions, the following four properties are incompatible:
\begin{enumerate}
\item Locality: $\widetilde D(p)$ is a smooth function of $p_\mu$ with the
period~$2\pi/a$.
\item No species doubling: $\widetilde D(p)$ is invertible for all $p\neq0$.
\item Skewness: $\widetilde D(-p)^T=-B\widetilde D(p)B^{-1}$.
\item Pseudo-parity invariance:
$[\widetilde D(p)+\widetilde D(-p)]^T=
+B[\widetilde D(p)+\widetilde D(-p)]B^{-1}$.
\end{enumerate}
\end{theorem}

We have $\widetilde D(-p)=-\widetilde D(p)$ from properties 3 and~4. By
substituting, say, $p_{\rm d}=(\pi/a,0,\cdots,0)$ into this relation, we
find that $\widetilde D(p_{\rm d})=\widetilde D(-p_{\rm d})=
-\widetilde D(p_{\rm d})=0$ from the periodicity stated in property~1. This is
in contradiction with property~2.~\endproof

\section{Discussion}

We must explain why a similar no-go theorem does not apply to dimensions other
than $d=8k$ and~$d=8k+1$. The Majorana decomposition in Euclidean field theory
is possible only for dimensions $d=0$, 1, 2, 3, $4\pmod8$
(See~Ref.~\citen{Inagaki:2004ar}).

In $8k+2$~dimensions, the skewness is expressed by $D(-z)^T=+BD(z)B^{-1}$, and,
corresponding to the pseudo-chiral invariance~(\ref{twoxeight}), we may
postulate~$\{\gamma,D(-z)\}^T=+B\{\gamma,D(z)\}B^{-1}$. The Wilson and overlap
Dirac operators possess this property due to the
relation~$D(-z)^T=+BD(z)B^{-1}$. However, a combination of these two properties
does {\it not\/} imply chiral invariance. In fact, it can be seen that these
two properties imply the same relation.

Similarly, in $8k+4$~dimensions, the skewness is expressed
by~$D(-z)^T=\break+B\gamma D(z)\gamma B^{-1}$, and the pseudo-chiral invariance
is replaced
by~$\{\gamma,D(-z)\}^T=+B\gamma\{\gamma,D(z)\}\gamma B^{-1}=
+B\{\gamma,D(z)\}B^{-1}$. The Wilson and overlap Dirac operators in fact
possess this property due to the relation~$D(-z)^T=+B\gamma D(z)\gamma B^{-1}$.
A combination of these two again does not imply chiral invariance, because,
again, these two imply the same relation.

For odd dimensions~$d=8k+3$, the skewness is expressed
by~$D(-z)^T=+BD(z)B^{-1}$. The pseudo-parity invariance~(\ref{twoxeleven}) is
replaced by~$[D(z)+D(-z)]^T=+B[D(z)+D(-z)]B^{-1}$. The Wilson--Dirac operator
possesses this property due to the relation~$D(-z)^T=+BD(z)B^{-1}$. The
overlap-Dirac operator does not have such a simple property and cannot be
utilized in these dimensions.\cite{Inagaki:2004ar} These two properties are,
however, the same and do not lead to parity invariance.

All the above facts are reflections of the properties of gamma matrices in each
dimension considered and are consistent with the fact that lattice Majorana
decomposition can be realized in dimensions other than $8k$
and~$8k+1$.\cite{Inagaki:2004ar}

In Ref.~\citen{Inagaki:2004ar}, it is argued that the difficulty involved in
the Majorana decomposition in $8k$ and $8k+1$ dimensions is a manifestation of
the global gauge anomaly in $8k$~dimensions. For this argument, the equivalence
of the Majorana fermion and the Weyl fermion in a real representation in
$8k$~dimensions, which holds in Minkowski spacetime and in the unregularized
Euclidean theory, is crucial. (A similar equivalence in $8k+4$~dimensions
holds even in Euclidean lattice gauge theory.\cite{Suzuki:2000ku}) In Euclidean
theories, the following correspondence is suggested:
\begin{equation}
   \chi(x)=\psi(x)+B^{-1}\overline\psi(x)^T,
\label{threexone}
\end{equation}
where $\psi(x)$ is the (left-handed) Weyl fermion:
\begin{equation}
   {1-\gamma\over2}\psi(x)=\psi(x),\qquad
   \overline\psi(x){1+\gamma\over2}=\overline\psi(x).
\end{equation}
If the expression~(\ref{threexone}) is substituted into the
action~(\ref{twoxone}) and {\it if\/} the pseudo-chiral
invariance~(\ref{twoxeight}) is assumed, the following action of the Weyl
fermion is obtained
\begin{equation}
   S_{\rm M}=a^d\sum_{x}a^d\sum_{y}\overline\psi(x)D(x-y)\psi(y).
\end{equation}
This is consistent with the conjectured equivalence of the Majorana fermion
and the Weyl fermion in $8k$~dimensions. As we have observed, however, the
above action suffers from species doubling (because we have assumed
pseudo-chiral invariance), and it cannot be utilized in this form.

On the basis of this conjectured equivalence of Majorana and Weyl fermions
in $8k$~dimensions, it might be thought possible to {\it define\/} a theory of
the Majorana fermion in $8k$~dimensions by using the lattice {\it Weyl\/}
fermion through~Eq.~(\ref{threexone}). (For a review of recent developments
regarding lattice Weyl fermions with an extensive list of references,
see~Ref.~\citen{Niedermayer:1998bi}.) In this approach, the partition function
of the Majorana fermion is given by the partition function of the Weyl fermion
and the correlation functions of the Majorana fermion are defined through
Eq.~(\ref{threexone}) from the correlation functions of the Weyl fermion. The
two-point function of the Weyl fermion is given by
\begin{equation}
   \langle\psi(x)\overline\psi(y)\rangle=\hat P_-{1\over D}P_+(x,y),
\end{equation}
where $\hat P_-=(1-\hat\gamma)/2$ and $\hat\gamma=\gamma(1-aD)$. The Dirac
operator~$D$ is assumed to satisfy the Ginsparg--Wilson
relation~$\gamma D+D\gamma=aD\gamma D$.\cite{Ginsparg:1982bj} By using
Eq.~(\ref{threexone}), we obtain
\begin{equation}
   \langle\chi(x)\chi(y)^T\rangle={1\over BD}(x,y)
   -{a\over2}B^{-1}a^{-d}\delta_{x,y},
\end{equation}
after some manipulation, or, taking the anti-symmetric part of the
right-hand side, we have
\begin{equation}
   \langle\chi(x)\chi(y)^T\rangle={1\over BD_A}(x,y).
\label{threexsix}
\end{equation}
If the overlap-Dirac operator is utilized as the Dirac operator~$D$, the
propagator~(\ref{threexsix}) acquires doubler's poles, as we have observed.
Thus, this natural approach based on the lattice Weyl fermion in the presently
considered form does not remove the difficulty.

We hope that our no-go theorem will be useful for investigating a possible
resolution of the difficulty concerning lattice Majorana fermions in $8k$ and
$8k+1$ dimensions.\footnote{For this purpose, it would be useful to note the
fact that, in $8k$ and~$8k+1$ dimensions, there exists a representation of the
Clifford algebra such that all the gamma matrices are real symmetric
and~$B=1$.} It seems rather non-trivial to avoid this difficulty involving
lattice Majorana fermions, due to its possible connection to the global gauge
anomaly. We should always keep in mind, however, that ``No-go theorems,
however, are frequently circumvented in an unexpected
way.''\cite{Hasenfratz:2004wy}

\section*{Acknowledgements}
I would like to thank Takanori Fujiwara for helpful discussions. A concise
proof of the Nielsen--Ninomiya theorem that I learned at Mini-workshop on
lattice field theory (the YITP workshop YITP-W-04-02) was quite helpful in
carrying out the present work. (This proof is reproduced in the appendix.) I
would like to thank the lecturer, Martin L\"uscher, for a series of lectures
and Yukawa Institute for Theoretical Physics at Kyoto University for
hospitality.

\appendix
\section{}
In this appendix, we present a concise proof of the Nielsen--Ninomiya theorem
for the reader's convenience. This is a short-cut version of the proof
presented in~Ref.~\citen{Friedan:nk}. The theorem is as follows.
\begin{theorem}
For the momentum representation of the free lattice Dirac operator
\begin{equation}
   \widetilde D(p)=a^d\sum_ze^{-ipz}D(z)
\end{equation}
in even $d$~dimensions, the following four properties are incompatible:
\begin{enumerate}
\item Locality: $\widetilde D(p)$ is a smooth function of $p_\mu$ with the
period~$2\pi/a$.
\item Correct dispersion: $\widetilde D(p)=i\sum_\mu\gamma_\mu p_\mu+O(ap^2)$.
\item No species doubling: $\widetilde D(p)$ is invertible for all $p\neq0$.
\item Chiral invariance: $\{\gamma,\widetilde D(p)\}=0$.
\end{enumerate}
\end{theorem}
From the momentum representation of the Dirac operator, we define the current
\begin{equation}
   j_\mu(p)=-{i^{d/2}\Gamma(d/2)\over2(2\pi)^{d/2}\Gamma(d)}
   \epsilon_{\mu\nu_1\cdots\nu_{d-1}}\tr\{
   \gamma[\widetilde D(p)\partial_{\nu_1}\widetilde D(p)^{-1}]
   \cdots[\widetilde D(p)\partial_{\nu_{d-1}}\widetilde D(p)^{-1}]\},
\end{equation}
where $\partial_\nu$ denotes the derivative with respect to the momentum:
$\partial_\nu\equiv\partial/(\partial p_\nu)$. This current~$j_\mu(p)$ is
well-defined for all $p\neq0$, from properties 1 and~3. For $p\neq0$, it can
further be verified that the current is conserved, i.e.,
$\partial_\mu j_\mu(p)=0$, by using property~4. Therefore, the surface integral
in the Brillouin zone
\begin{equation}
   I=\int_S\rmd\sigma_\mu\,j_\mu(p),
\end{equation}
is independent of the surface~$S$, as long as $S$ surrounds the origin, $p=0$.
This integral can therefore be evaluated over an infinitesimally small
sphere around the origin by using the asymptotic form given in property~2. We
then obtain~$I=1$. On the other hand, if we take the boundary of the Brillouin
zone as the surface~$S$, we have~$I=0$ from the periodicity in property~1. This
is a contradiction.\endproof

\end{document}